\documentclass[aps,prl,reprint]{revtex4-1}
\usepackage[english]{babel}
\usepackage[utf8]{inputenc}
\usepackage{xcolor}
\usepackage{physics}
\usepackage[colorinlistoftodos, color=green!40, prependcaption]{todonotes}
\usepackage{amsthm,amsmath,amssymb}
\usepackage{mathtools}
\usepackage{physics}
\usepackage{graphicx}
\usepackage[left=15mm,right=15mm,top=20mm,columnsep=15pt]{geometry} 
\usepackage{adjustbox}
\usepackage{placeins}
\usepackage[T1]{fontenc}
\usepackage{lipsum}
\usepackage{csquotes}
\usepackage{float}
\usepackage[pdftex, pdftitle={Article}, pdfauthor={Author}]{hyperref} 

\def\bs#1{\boldsymbol{#1}}
\def\rmd{{\rm d}}

\newcommand{\markblue}[1]{\textcolor{black}{#1}}

\begin{document}

\title{Single-particle diffraction with a hydrodynamic pilot-wave model}

 \author{Giuseppe Pucci$^{1,2,3,*}$, Antoine Bellaigue$^{2}$, Alessia Cirimele$^{4}$, Giuseppe Alì$^{4}$, Anand U. Oza$^{5}$}

 \email[Correspondence email addresses: ]{giuseppe.pucci@cnr.it, oza@njit.edu}

 \affiliation{$^{1}$Consiglio Nazionale delle Ricerche - Istituto di Nanotecnologia (CNR-Nanotec), Via P. Bucci 33C, 87036 Rende (CS), Italy,}
  
 \affiliation{$^{2}$Univ Rennes, CNRS, IPR (Institut de Physique de Rennes)—UMR 6251, F-35000 Rennes, France,}

 \affiliation{$^{3}$INFN, Sezione di Lecce, Via per Monteroni, Lecce 73100, Italy,}
 
 \affiliation{$^{4}$Dipartimento di Fisica, Universit\`a della Calabria, Via P. Bucci 31C, 87036 Rende (CS), Italy,}
  
 \affiliation{$^{5}$Department of Mathematical Sciences \& Center for Applied Mathematics and Statistics, New Jersey Institute of Technology, Newark, New Jersey 07102, USA}

\date{\today} 

\begin{abstract}

A macroscopic hydrodynamic system that couples a particle and a wave has recently renewed interest in the question as to what extent a classical system may reproduce quantum phenomena. Here we investigate single-particle diffraction with a pilot-wave model originally developed to describe the hydrodynamic system. We study single-particle interactions with a barrier and slits of increasing width by focusing on the near field. We find single-particle diffraction arising as wavelike patterns in the particles' position statistics, which we compare to the predictions of quantum mechanics. We provide a mechanism that rationalizes the diffractive behavior in our system.


\end{abstract}

\maketitle


Diffraction is a phenomenon that attracts great interest for its implications in fundamental physics \cite{feynman_feynman_1966} and wide range of applications \cite{bunaciu2015x,filippetto2022ultrafast}. 
In classical physics, diffraction is fully captured as a wave phenomenon that occurs when a wave meets an obstacle or an aperture in a screen \cite{born2013principles}. 
The possibility of single-particle diffraction~\cite{Taylor1909} first arose with the discovery of light quanta~\cite{Planck,Einstein}. 
The wave-like nature of massive particles was postulated shortly thereafter~\cite{de1924recherches}, and has been observed in landmark experiments on the diffraction of electrons~\cite{Merli1976,tonomura_demonstration_1989,bach_controlled_2013}. 


Over the last two decades, a classical macroscopic system that couples a particle and a wave \cite{Couder2005a} has shown a number of analogies with quantum particles \cite{bush_hydrodynamic_2021,Bush2024}, including quantized orbital radius~\cite{Fort2010,Harris2014a,blitstein2024minimal} and angular momentum \cite{perrard_self-organization_2014,durey_faraday_2017}, Zeeman-like splitting~\cite{eddi_level_2012}, wavelike statistics \cite{harris_wavelike_2013,Gilet2014,gilet_quantumlike_2016,Saenz2018,sungar2021walking} \markblue{and statistical projection}~\cite{Saenz2018} in cavities, tunneling~\cite{Eddi2009b,Hubert2017,Nachbin2017,Tadrist2019}, Friedel oscillations~\cite{Saenz2019a}, \markblue{spin lattices~\cite{saenz_emergent_2021}, surreal trajectories~\cite{Frumkin2022}, interaction-free measurement~\cite{Frumkin2023a}, superradiance effects~\cite{papatryfonos2022hydrodynamic,Frumkin2023b}, Anderson localization~\cite{abraham2024anderson}, static Bell test \cite{papatryfonos2024static} and other pair correlations~\cite{Nachbin2018,Valani2018_2,nachbin2022effect}}. In this system, the particle is a droplet coupled to the wave field that it generates by bouncing on the surface of a vibrating liquid bath \cite{Couder2005a,protiere_particlewave_2006,Molacek2013b}. The ensemble of drop and surface wave field has been termed a {\it walker}. 

Seminal experiments and simulations with walkers \markblue{reported} single-particle diffraction when the walker crossed an aperture between two submerged barriers~\cite{couder_single-particle_2006,couder2012probabilities}. While these studies prompted investigations of diffractive behavior in other classical systems~\cite{Schiebel2019,Rieser2019}, they were questioned on the basis of an insufficient number of data points \cite{Andersen2015,Bohr2016}, and later experiments with finer control of experimental parameters 
could not reproduce the original results~\cite{pucci_walking_2018,rode2019,ellegaard_interaction_2020,Ellegaard2024}. While these later experiments did not exclude the possibility of obtaining statistical distributions similar to the first experiments 
in some corner of parameter space, \markblue{the distributions were quite different from those of quantum particles.} 
\markblue{Indeed, these works demonstrated that observing quantum-like diffraction of walkers in experiments is very difficult} since the barriers are submerged under a relatively thin layer of liquid, which causes specific particle-barrier interactions \cite{Faria2017,pucci_walking_2018,ellegaard_interaction_2020,Ellegaard2024}. On the theoretical side, a Green's-function model of the interaction of walkers with an aperture yielded small scattering angles, so the associated statistical distributions were different from both the walker system and quantum particles~\cite{Dubertrand2016}. Despite the large amount of work on this subject, it has remained an open question as \markblue{to whether quantum-like single-particle diffraction can be observed in the walker system.} 

\markblue{We here answer this question affirmatively by generalizing the so-called {\it stroboscopic model} of walkers \cite{oza_trajectory_2013}. Previous work demonstrated that this hydrodynamic pilot-wave model}
successfully reproduces analogs of two canonical quantum phenomena, namely, the quantized states exhibited by particles in a rotating frame~\cite{Oza2014a,Oza2014b} and in a simple harmonic potential~\cite{Labousse2016a,Kurianski2017}, as well as many other related systems~\cite{Valani2022,Perrard2018,Budanur2019,Montes2021,Tambasco2018a,Valani2018,Couchman2020,Thomson2019,Thomson2020,Thomson2021,Barnes2020,Couchman2022, saenz_emergent_2021}. While the model exhibits excellent agreement with experiments in unbounded geometries~\cite{Oza2014a,Oza2014b,Labousse2016a,Thomson2019,Couchman2019,Couchman2020}, it has not yet been used to model the interactions of particles with barriers.
\begin{figure*}[hbtp]
    \centering
    \includegraphics[width=1\textwidth]{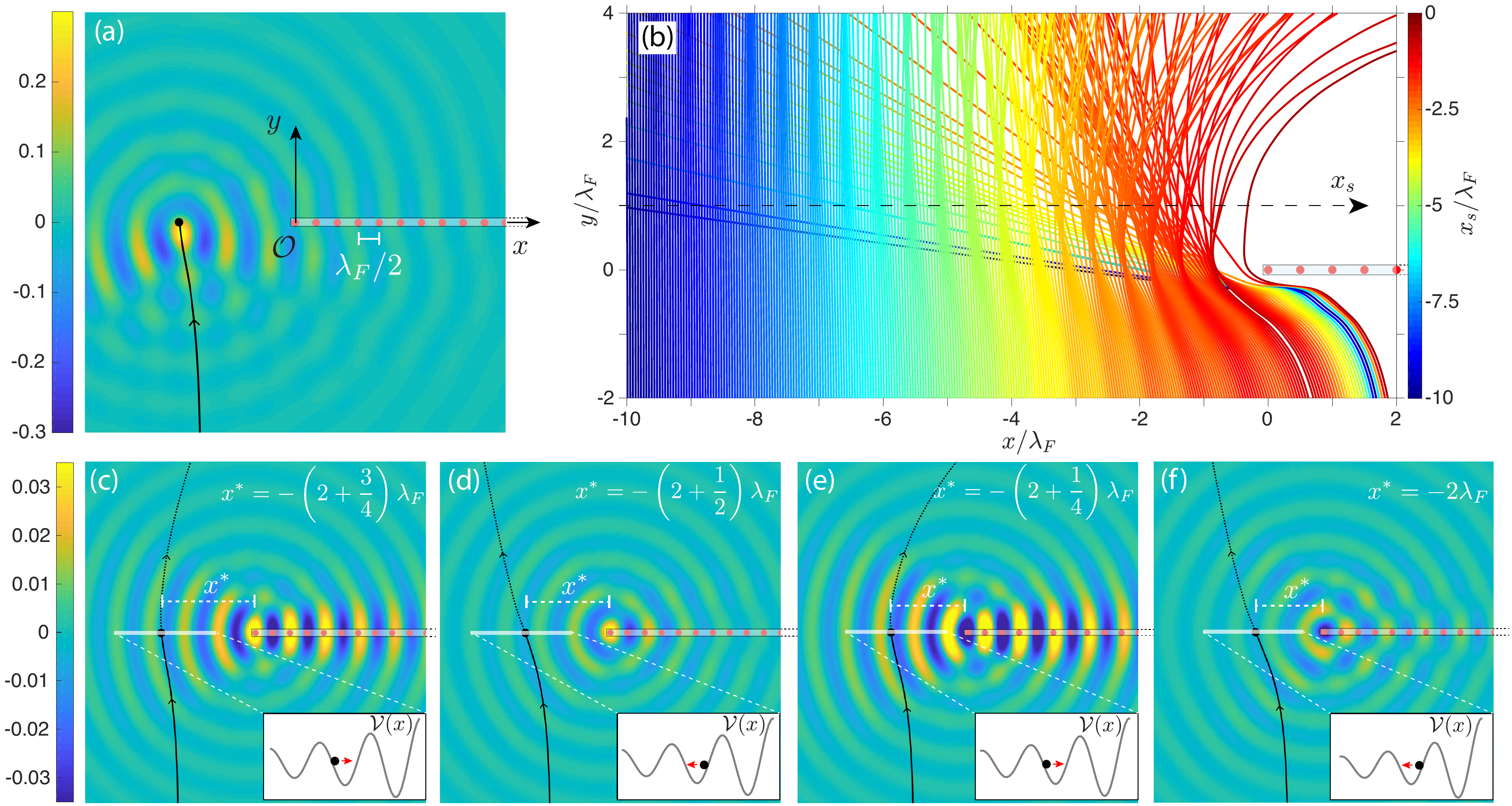}
    \caption{Simulation of walkers interacting with a linear barrier made of a series of secondary sources. (a) Snapshot of a trajectory with the full wave field $\mathcal{H}(\bs{x},t)$; see also Supplementary Movie 1. (b) Selection of trajectories color-coded by their position $x_s$ on a screen located at $y=\lambda_F$, which highlights the self-focusing mechanism. 
    (c--f) Wavefields $\mathcal{H}_s(\bs{x},t^*)$ due to the secondary sources alone for four trajectories, with insets showing the effective source potential $\mathcal{V}(x)$ (not to scale); see also Supplementary Movie 2. Colorbars are in arbitrary units.}
    \label{fig:1}
\end{figure*}

We here model the interaction of single walkers with linear barriers described by series of secondary sources, similarly to the pioneering simulations on this subject \cite{couder_single-particle_2006} but using the stroboscopic model \cite{oza_trajectory_2013}. 
\markblue{While} previous studies have searched for \markblue{quantum-like diffractive} behavior in the {\it far field}
~\cite{couder_single-particle_2006,couder2012probabilities,andersen_double-slit_2015,Bohr2016,Dubertrand2016,pucci_walking_2018,ellegaard_interaction_2020,Ellegaard2024}, we \markblue{find it in the distribution of particle positions in} 
the {\it near field}, that is, a few wavelengths away from the barriers. 
In the single-slit geometry, the number of peaks in the distribution increases with the slit's width, \markblue{in agreement with quantum mechanics}. We rationalize the emergence of the distributions in terms of the wavefield generated by the secondary sources, which behaves as an effective transient potential for the particle. We provide a detailed comparison with diffraction in quantum mechanics. The distributions are qualitatively similar, with the periodicity of the walker's distribution being half that of quantum particles. 


\begin{figure}[hbtp]
    \centering
    \includegraphics[width=1\columnwidth]{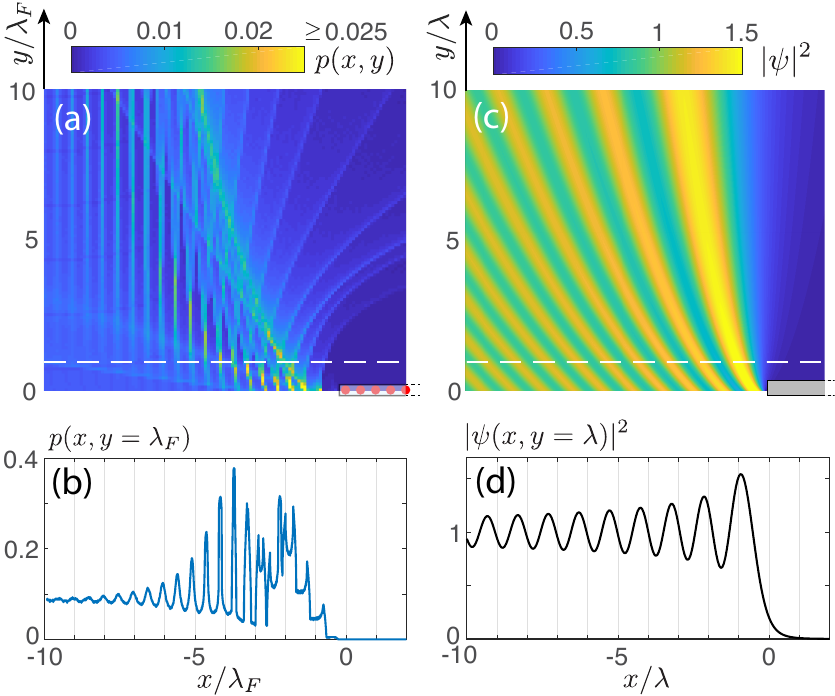}
    \caption{Comparison between the probability density functions (PDFs) of walkers (a,b) and quantum particles (c,d) interacting with an edge. 
    (a) PDF $p(x,y)$ of walkers. (b) PDF of walker impacts on a screen placed at a distance $\lambda_F$ from the edge (white dashed line in (a)). (c) Unnormalized probability density $|\psi|^2$ of quantum particles. 
     (d) Variation in $|\psi|^2$ at a distance $\lambda$ from the edge, where $\lambda$ is the de Broglie wavelength. 
    }
    \label{fig:2}
\end{figure}

\begin{figure*}[hbtp]
    \centering
    \includegraphics[width=1\textwidth]{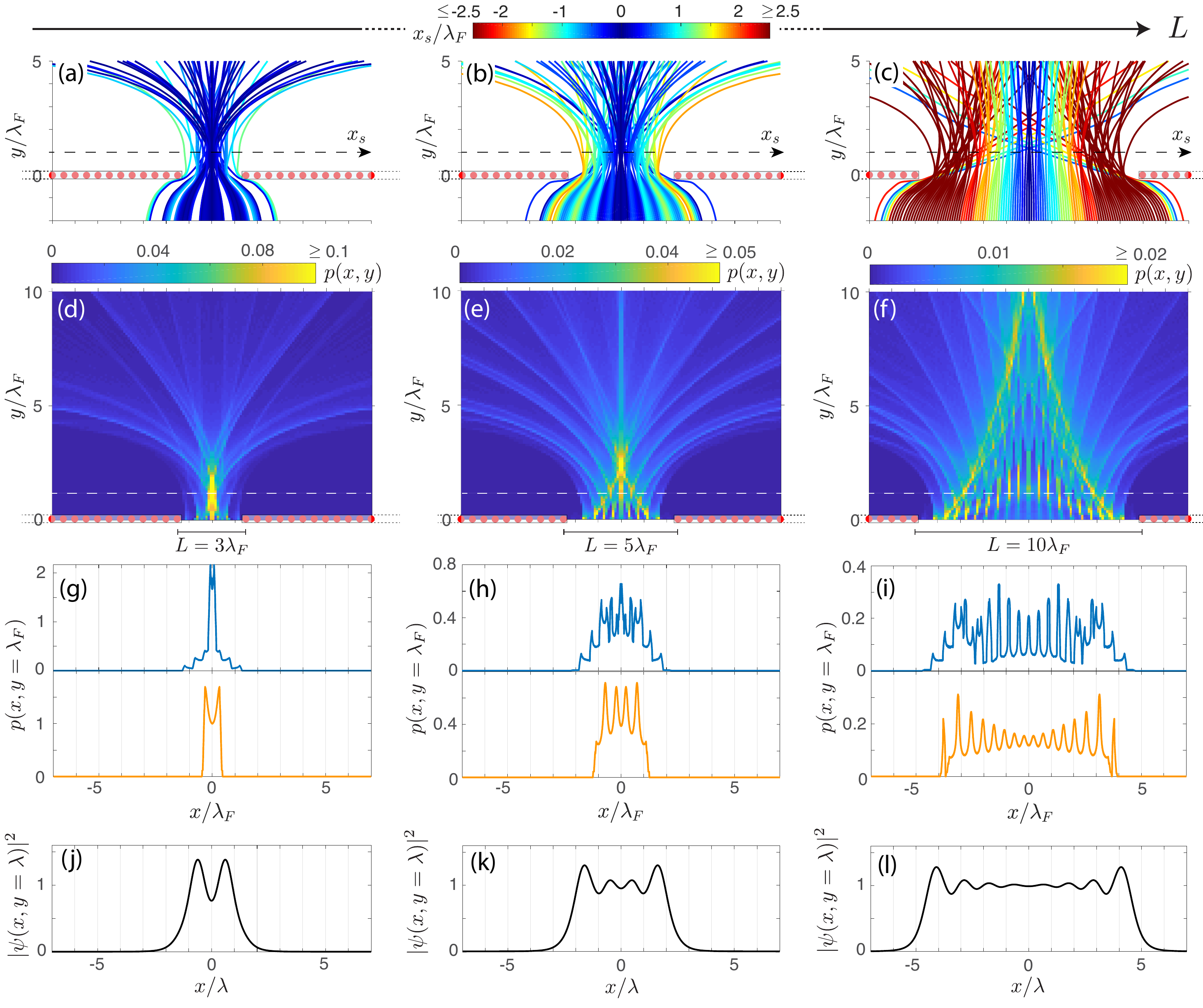}
    \caption{Diffraction of walkers past a single slit for three different slit widths: $L = 3\lambda_F$ (a,d,g); $L = 5\lambda_F$ (b,e,h) and $L = 10\lambda_F$ (c,f,i); see also Supplementary Movie 3. (a--c) Trajectories are color-coded by their impact position $x_s$ on a screen located at $y=\lambda_F$ (dashed line). (d--f) PDFs of walkers. (g--i) PDFs of walker impacts on the screen. \markblue{Top (bottom) panels correspond to the secondary source spacing $\Delta x/\lambda_F = 0.5$ (0.3).} (j--l) Variation in $|\psi|^2$ for quantum particles at a distance $\lambda$ from the slit.}
    \label{fig:3}
\end{figure*}

{\it Model description.}\textemdash We consider two geometries: (1) a semi-infinite linear barrier, or an {\it edge}, as shown in Fig.~\ref{fig:1}, and (2) two semi-infinite linear barriers separated by a distance $L$, or a {\it single-slit}, as shown in Fig.~\ref{fig:3}. 
Both geometries are represented by linear arrays of pointlike sources separated by a distance $\Delta x$, specifically, at $\bs{x}_j=(j\Delta x,0)$ for $j\in\mathbb{N}$ for the edge [Fig.~\ref{fig:1}(a)] and $\bs{x}_j=\pm(L/2+j\Delta x,0)$ for the single slit [Fig.~\ref{fig:3}]. \markblue{Unless otherwise stated, the results presented herein are for $\Delta x = \lambda_F/2$, where $\lambda_F$ is the wavelength of the subthreshold Faraday waves generated by the walker~\cite{Faraday1831}. 
We find that the statistical behavior is qualitatively independent of $\Delta x$, the effect of which is assessed analytically using a simplified model in~\cite[\S III]{Supplement} and numerically using the full model in~\cite[\S VIII]{Supplement}}. In a typical walker experiment, a drop of mass $m$ with horizontal position $\bs{X}_p(t)=(X_p(t),Y_p(t))$ bounces periodically with period $T_F = 2/f$ in the presence of a gravitational acceleration $g$ on the surface of a fluid bath vibrating with frequency $f$ and peak acceleration $\gamma < \gamma_F$, $\gamma_F$ being the so-called Faraday instability threshold~\cite{Faraday1831}. The stroboscopic model reads~\cite{oza_trajectory_2013} 
\begin{align}
m\ddot{\bs X}_p+D\dot{\bs X}_p&=-mg\bs\nabla \mathcal{H}(\bs X_p,t),\nonumber \\
\text{where }\mathcal{H}(\bs{x},t)&=\frac{1}{T_F}\int_{-\infty}^t h(\bs{x},s)\mathrm{e}^{-(t-s)/T_M}\,\rmd s\label{TrajEqn}
\end{align}
and $T_M$ is the ``memory'' timescale over which the waves generated by the walker decay~\cite{Eddi2011a,Molacek2013b}. That is, the drop moves in response to two horizontal forces: a wave force $-mg\bs{\nabla} \mathcal{H}(\bs{X}_p,t)$ proportional to the local slope of the wave height $\mathcal{H}$, and a drag $-D\dot{\bs{X}}_p$ experienced during impact and flight~\cite{Molacek2013b}.
To model the wave field, we first assume that, in the absence of barriers (free space), the impact of a walker at the origin generates a standing wave $h_0(r)$ that is monochromatic of wavelength $\lambda_F$ in the near field and decays exponentially over a lengthscale $d$~\cite[Eq.~(1)]{Supplement}. The wave profile due to a droplet impact at $\bs{X}_p(t)$ at time $t$ in the presence of barriers is thus
\begin{align}
h(\bs{x},t)&=h_0(|\bs{x}-\bs{X}_p(t)|)+h_s(\bs{x},t),\label{Wave}
\end{align}
where the wave field $h_s(\bs{x},t)$ due to the barriers is determined by imposing the Dirichlet boundary condition  
 \begin{align}
 h(\bs{x}_j,t)=0 
 \label{DirBC}
 \end{align}
on each barrier; 
that is, the wave field is zero for all time on the secondary sources. 
The constants $D$, $\lambda_F$, $d$ and $T_M$ are known in terms of fluid parameters~\cite[\S I.A]{Supplement}. We note that, while a linear array of secondary sources separated by $\lambda_F/2$ was also used in the original simulations of Couder \& Fort~\cite{couder_single-particle_2006}, they did not enforce the boundary conditions~\eqref{DirBC} exactly. We simulate $10^4$ trajectories per geometry, discarding those that cross the barrier~\cite[Table I]{Supplement},
and report here the results obtained for $\gamma/\gamma_F=0.95$, for which $d\approx 3\lambda_F$. Results for lower and higher forcings are reported in the Supplementary Material~\cite[\S V]{Supplement}.   



{\it Wave-like diffraction from an edge.}\textemdash In our first simulations, particles are launched one-by-one with normal incidence at an edge, as shown in Fig.~\ref{fig:1}(a,b). 
We observe that the trajectories focus towards certain regions of space, while other regions of space are relatively devoid of trajectories. This self-focusing mechanism is evident in Fig.~\ref{fig:1}(b). In the vicinity of the edge, some of these trajectories are deflected to the left away from the edge, while others are deflected behind the edge. 

To obtain physical insight into the self-focusing mechanism, we consider the wave field $\mathcal{H}_s(\bs{x},t)\equiv \int_{-\infty}^th_s(\bs{x},s)\mathrm{exp}[-(t-s)/T_M]\,\rmd s$ due to the secondary sources for four different trajectories. We note from Eq.~\eqref{TrajEqn} that $\mathcal{H}_s$ plays the role of a potential energy due to the secondary sources. $\mathcal{H}_s(\bs{x},t)$ reaches the highest amplitude and momentarily ``freezes\rq\rq\, when the drop crosses the plane of the edge, $\bs{X}_p(t^*)\equiv (x^*,0)$ (Supplementary Movie 2). 
This critical time $t=t^*$ is when the waves emitted by the sources most constructively interfere and are thus responsible for most of the walker's deflection \cite[\S II]{Supplement}. From Figs.~\ref{fig:1}(c--f) we observe that the walkers are deflected by the local gradient in $\mathcal{H}_s(\bs{x},t^*)$, with two trajectories being deflected leftward [Fig.~\ref{fig:1}(d,f)] and the other two rightward [Fig.~\ref{fig:1}(c,e)]. The insets, which show the effective source potential $\mathcal{V}(x)\equiv \mathcal{H}_s((x,0),t^*)$, serve to further highlight this mechanism, as each walker is pushed laterally toward the nearest trough in $\mathcal{V}(x)$.

The trajectories can be used to construct a probability density function (PDF) $p(x,y)$ of walker positions (Fig.~\ref{fig:2}(a,b)). 
As a comparison, we show in Fig.~\ref{fig:2}(c,d) the 
solution 
for the diffraction of a quantum mechanical 
plane wave $\psi$ by an edge in 2D, as obtained by solving Schr\"{o}dinger's equation with the appropriate boundary conditions~\cite[\S IV]{Supplement}.
There are qualitative similarities between the two systems: both exhibit spatial oscillations in amplitude, which in the walker system reflect the self-focusing of trajectories observed in Fig.~\ref{fig:1}(b). There is a pronounced bright beam in both systems that emanates from the edge and curves leftward [Fig.~\ref{fig:2}(a,c)], which in the walker system indicates a relatively large density of walkers which cross that particular path. Both systems also have a small (but nonzero) amplitude shadow region behind the edge. 
The 1D position PDFs [Fig.~\ref{fig:2}(b,d)] of both the walker and quantum systems exhibit relatively large amplitudes near the edge, 
with oscillations that decay farther away from the edge.  
The oscillations decay algebraically in the quantum system but exponentially in the walker system because the latter is dissipative and thus has spatially damped waves. 
We observe that the quantum probability density oscillates on the de Broglie wavelength $\lambda$ [Fig.~\ref{fig:2}(d)], while the walker screen position PDF oscillates on {\it half} the Faraday wavelength, $\lambda_F/2$ [Fig.~\ref{fig:2}(b)]. The half-wavelength periodicity of the walker statistics may be intuited by observing that the secondary source wavefield $\mathcal{H}_s(\bs{x},t^*)$ flips sign as $x^*$ is increased by $\lambda_F/2$ [Fig.~\ref{fig:1}(c,e)], suggesting that the source potential $\mathcal{V}(x^*)$ is $\lambda_F/2$-periodic in $x^*$ (see ~\cite[\S III]{Supplement} for a quantitative explanation). We note that in the quantum system the regions of relatively large intensity [yellow regions in Fig.~\ref{fig:2}(c)] curve to the left, while they are mostly straight upward in the walker system. 


{\it Wave-like diffraction by a single slit.}\textemdash In the next set of simulations, walkers were launched one-by-one with normal incidence towards two linear barriers separated by a distance $L$, as depicted in Fig.~\ref{fig:3}(a--c). 
The self-focusing mechanism present for the edge [Fig.~\ref{fig:1}] is also evident for the single slit [Fig.~\ref{fig:3}]. Trajectories near the center are funneled into preferred regions of space, as evidenced by the yellow streaks in the 2D PDFs [Fig.~\ref{fig:3}(d--f)], while trajectories with impact parameters near the barriers are strongly deflected. The PDFs of screen impact positions $x_s$ [Fig.~\ref{fig:3}(g--i)] may be compared with their counterparts from quantum mechanics [Fig.~\ref{fig:3}(j--l)], 
the most salient feature being that the number of oscillations in the PDF increases with the slit width $L$. \markblue{This feature is independent of the source spacing $\Delta x$ in the walker system, as is evident by comparing the top and bottom panels in Fig.~\ref{fig:3}(g--i).} A mechanism for this phenomenon may be inferred from the trajectories in Fig.~\ref{fig:3}(a--c), as their $x_s$--values oscillate within a larger range as $L$ is increased progressively. 
Moreover, as the screen is placed farther from the barriers, the oscillations in the central region of the PDF are suppressed in both amplitude and lateral extent, 
in agreement with the predictions of quantum mechanics~\cite[Fig. 6]{Supplement}. 
There are again some differences between the walker and quantum systems; most notably, the walker PDFs exhibit oscillations on $\lambda_F/2$ while the quantum intensity oscillates on $\lambda$, a behavior that may be rationalized using the argument presented for the edge geometry in the preceding section. 

{\it Discussion.}\textemdash We have used a hydrodynamic pilot-wave model to investigate single-particle diffraction by linear barriers. 
Because the system is dissipative and thus the walker wavefield decays exponentially in space, we focused on the near field, that is, a few wavelengths away from the barriers, a regime that was not explored in prior studies~\cite{couder_single-particle_2006,couder2012probabilities,andersen_double-slit_2015,Bohr2016,Dubertrand2016,pucci_walking_2018,ellegaard_interaction_2020,Ellegaard2024}. We found wavelike particle statistics that exhibit oscillations on half the wavelength of the walker wave field, that is, $\lambda_F/2$. We have rationalized this behavior in terms of the wave field generated by the barriers, which creates an effective transient potential responsible for walker deflection and self-focusing. 
We have compared our results to the diffraction of a plane wave in the same geometries as described by  
quantum theory. The distributions are qualitatively similar but differ in their periodicity, which for quantum particles is generally on the order of one de Broglie wavelength. 

\markblue{The self-focusing mechanism that we propose is reminiscent of that responsible for the quantization of a walker's quasi-periodic orbits~\cite{Fort2010,Perrard2014,Labousse2014a,Labousse2016a}. The key difference is that the effective potential $\mathcal{H}_s(\bs{x},t)$ in our system is transient, dominating at the time $t=t^*$ when the walker passes through the plane of the barrier, while the potential in the orbital case is quasi-stationary because the walker is effectively confined in space. Prior work has demonstrated that the stationary points of the curvature of the droplet's trajectory are responsible for quantum-like behavior~\cite{blitstein2024minimal}; while these stationary points are absent in our system, the secondary sources play an analogous role, as they are momentarily frozen at $t=t^*$ (Supplementary Movie 2,~\cite[\S II]{Supplement}). Furthermore, the transient potential in our system is reminiscent of the standing wavefield responsible for Friedel-like oscillations when a walker interacts with a submerged well~\cite{Saenz2019a}. While quantum-like probability distributions in that system are due to oscillations in the walker speed, here they are due to the aforementioned self-focusing mechanism.} We also note that the diffractive behavior in our system arises from a non-chaotic dynamics, in which the position at which the walker hits a screen generally has a predictable dependence on the initial conditions~\cite[Figs. 7 and 8]{Supplement}. 

Our results establish \markblue{near-field} quantum-like diffraction with a classical wave-driven system. 
The results presented here can inspire future works on single-particle diffraction with generalized \cite{bush_hydrodynamic_2021,Durey2020} or walker-inspired pilot-wave theories \cite{Durey2020b,Durey2020c,Dagan2020,darrow2024revisiting}, which could yield results even closer to quantum mechanics. 


\begin{acknowledgements}
A. O. acknowledges support from NSF DMS-2108839. G. P. acknowledges the CNRS Momentum program, and thanks Pedro Sáenz for fruitful discussions.
\end{acknowledgements}

\bibliographystyle{apsrev4-2}
\bibliography{biblio,ARFMBib}

\end{document}